\title{ORDERED STATES  \\ IN THE \\ DISORDERED HUBBARD MODEL}
\author{ 
P.~J.~H.~Denteneer\thanks{e-mail address: pjhdent@lorentz.leidenuniv.nl} 
      \\  {\em Instituut--Lorentz, University of Leiden,} \\
          {\em P. O. Box 9506, 2300 RA Leiden, The Netherlands, } \\~\\
M.~Ulmke \\ {\em Theoretische Physik III, Universit\"at Augsburg,} \\ 
            {\em D 86135 Augsburg, Germany} \\~\\
R.~T.~Scalettar, G.~T.~Zimanyi \\
            {\em Department of Physics, University of California,} \\
            {\em Davis, CA 95616, USA} }

\date{}
\documentstyle[12pt,psfig]{article}
\newcommand{\be}{\begin{equation}}
\newcommand{\ee}{\end{equation}}
\newcommand{\bc}{\begin{center}}
\newcommand{\ec}{\end{center}}
\newcommand{\ha}{{\scriptstyle \frac{1}{2}}}
\newcommand{\eps}{\varepsilon}
\newcommand{\etal}{{\em et al.}}

\begin{document}
\hoffset=-0.0 true cm
\voffset=-2.0 true cm
\maketitle
\vspace{1cm}
\begin{abstract}
The Hubbard model is studied in which disorder is introduced by putting
the on-site interaction to zero on a fraction $f$ of 
(impurity) sites of a square lattice. Using Quantum Monte Carlo
methods and Dynamical Mean--Field Theory we find that 
antiferromagnetic long--range order is initially enhanced 
at half--filling and stabilized off half-filling by the disorder. 
The Mott-Hubbard charge gap of the pure system is broken
up in two pieces by the disorder: one incompressible state remains at 
average density $n=1$ and another can be seen slightly below $n=1+f$.
Qualitative explanations are provided.
\end{abstract}
\vspace{1.0cm}
Short title: Order in the disordered Hubbard model 

\newpage
 
\section{Introduction}
The problem of the interplay between disorder and interactions in 
systems of electrons is challenging and has a long history \cite{RMPS}.
Disorder can, by itself, localize electrons and make the conductivity
disappear, the {\em Anderson transition}. At appropriate densities, 
interactions can also cause the formation of insulating states via the
{\em Mott transition}. In the latter case, the insulating
states often possess additional magnetic or charge ordering. 
The simultaneous presence of disorder and interactions is
realized in many systems in nature, with fundamental
and fascinating consequences, e.g. the unusual
magnetic properties of heavily doped, compensated
semiconductors, and the unique transport properties of
two--dimensional and bulk superconductors \cite{RMPS}.
If both disorder and interactions are present, the
effects can sometimes reinforce each other, but can also compete.
In classical spin models, it has recently become clear that quenched
disorder can turn (temperature-driven) symmetry-breaking first order
phase transitions into continuous phase transitions,
possibly with new intervening critical points \cite{Berker}.
Although the arguments do not appear to apply directly to interacting
fermions, disorder may lead to similarly interesting phenomena.

The Hubbard model for interacting electrons on a lattice
\cite{Hub}
exhibits both Mott metal--insulator and magnetic 
phase transitions.  
On the one hand, it is expected that the interaction $U$
induces a gap at half--filling 
by separating many--body states with doubly
occupied orbitals from those with holes, when the on-site repulsion 
becomes larger than the non--interacting bandwidth. 
On the other hand, also near half--filling,
there is a tendency to antiferromagnetic (AF) ordering 
of the electron spins.  One could now ask the question
whether
the Mott transition in fact ever occurs in the absence of some
associated symmetry breaking such as magnetic order \cite{Gebhard97}.            
For interacting {\em bosons} the Mott transition (in this case from a 
superfluid) to an insulator is known to occur without any other 
accompanying order parameter \cite{dirtbos}.

To investigate the behavior of disordered interacting electron systems 
it is of interest to consider the
Hubbard Hamiltonian into which disorder is introduced: \\
\begin{equation}
{\hat H}_{\rm H} = - \sum_{i,j,\sigma } t_{ij} c_{i\sigma}^{\dagger} 
c_{j\sigma}^{\phantom \dagger}
+ \sum_{j} U_j (n_{j \uparrow}-\ha)(n_{j \downarrow}-\ha) 
-  \sum_{j,\sigma} (\mu - \eps_j) n_{j \sigma} , \label{eq:Hhdis}
\end{equation}
where $c_{j\sigma}$ is the 
annihilation operator for an electron at
site $j$ with spin $\sigma$.  
$t_{ij}$ is the hopping integral (non--zero only between neighboring 
sites $i$ and $j$),
$U_j$ the on-site interaction at site $j$ between electrons of 
opposite spin, and 
$n_{j \sigma}=c_{j \sigma}^{\dagger}c_{j \sigma}^{\phantom \dagger}$ 
is the occupation number operator. Various kinds of disorder are
allowed for in (\ref{eq:Hhdis}): (i) in the hopping integrals $t_{ij}$,
(ii) in the on--site {\em energy} $\eps_j$, or 
(iii) in the on--site {\em interaction} $U_j$.
Traditionally, because the case of non-interacting electrons was
considered first, attention has been devoted to the first two kinds of
disorder, termed {\em off-diagonal} and {\em diagonal disorder},
respectively.
In recent Quantum Monte Carlo simulations, the first two kinds 
were studied including {\em uniform} interactions, 
showing the destruction of AF 
long--range order at half--filling
for some critical strength of disorder in case (i), and allowing for
few conclusions in case (ii) because of sign problems already at
half--filling \cite{UlmRTS}. In the present paper, we study case (iii), i.e.
the Hamiltonian:
\be
{\hat H}  = - t \sum_{\langle i,j \rangle,\sigma } 
c_{i\sigma}^{\dagger} c_{j \sigma}^{\phantom \dagger}
+ \sum_{j} U_j (n_{j \uparrow}-\ha)(n_{j \downarrow}-\ha) 
-  \mu \sum_{j,\sigma} n_{j \sigma} , \label{eq:Hh}
\ee
in which  
the on-site repulsion $U_j$ is turned off from $U$ on a fraction $f$ 
of all sites ($\langle i,j \rangle$ denotes neighboring sites $i$ and $j$), 
i.e. the $U_j$ are taken from a bimodal distribution, 
\be
P \left( U_j \right) = (1-f) \, \delta \left( U_j - U \right) + 
                 f \, \delta \left( U_j \right) ~. \label{eq:bimod}
\ee
A physical realization of this kind of disorder could be 
impurity atoms in a crystal which can accomodate two electrons 
with a strongly reduced Coulomb repulsion, e.g. non-magnetic impurities
(like Zn) replacing copper atoms in the copper oxides.
Here, we do not want to attach ourselves to a specific physical
system, but try to understand what happens to ordered states of
interacting electrons when disorder of kind (iii) is introduced.
One expects the Mott transition to be shifted to average density 
$n = 1+f$ (at which density also the $U \neq 0$ sites start to be
doubly occupied), so that it is perhaps separated from the 
AF magnetic order which is likely to stay at $n = 1$. New types of
ordered states may also result.

To explore these phenomena tractable and reliable theoretical methods 
are hard to come by.
Specifically, mean--field approaches 
to the physics of strongly correlated quantum systems,
like the Hartree--Fock 
approximation and the slave--boson mean--field approach, 
can provide significant insight into the possible phases which
can arise due to interaction effects, but are well--known
to overestimate significantly
the tendency to form ordered states \cite{PDSBMF,GZEA}.
Other analytical approaches like the renormalization group,
while powerful, likewise have their limitations \cite{RMPS}.
The most obvious numerical approach, 
exact diagonalization of the Hamiltonian,
is restricted to small lattices (10--20 particles) \cite{Dagot}.
In this paper, we intend to study 
the effects of interactions and 
randomness 
by means of a stochastic sampling of the state space,
the Quantum Monte Carlo (QMC) method.
In this approach, correlation effects are treated exactly.
A few hundred interacting electrons can be simulated, roughly
an order of magnitude larger than by exact diagonalization.
While the constraints of this finite size are still very
significant, this number of particles is often
sufficient to allow systematic extrapolations
to bulk behavior, at least in two dimensions.
The QMC method we use is the {\em Determinant} Monte Carlo method
\cite{BlnkWh}, which has been
extensively applied to non--random interacting fermion Hamiltonians.
We combine these simulations 
with calculations using Dynamical Mean-Field Theory (DMFT), or: the 
{\it infinite--dimension} approach \cite{MV}. 
This technique improves significantly upon the Hartree--Fock approximation
since it is non--perturbative and takes quantum fluctuations into account
via a dynamical self-energy.
The infinite--dimension approach allows us to analyze problems in
the thermodynamic limit and in
larger ranges of parameter space, e.g. lower temperatures,
than accessible to the determinant algorithm.
DMFT has provided valuable 
insight into the physics of strongly correlated electron systems, 
e.g. the Mott--Hubbard transition \cite{Gebhard97,MV}.
Both techniques, which recently have been applied to
disordered Hubbard models also \cite{UlmRTS,DobKot,UlmJan},
will be briefly described in Section 2. In subsequent sections, we
introduce the quantities of interest that are calculated, present
results and discuss them.

\section{Quantum Monte Carlo and Dynamical Mean Field Theory}

\subsection{Determinant Monte Carlo method} \label{DetQMC}
A brief description of the Determinant Monte Carlo algorithm
for the Hubbard Hamiltonian (\ref{eq:Hh}) is given.  
One is interested in
computing operator expectation values like:
\be
\langle \hat A \rangle = 
{ {\rm Tr}\,\hat A e^{-\beta \hat H} \over {\rm Tr} \, 
e^{-\beta \hat H} } ~.  \label{eq:ave}
\ee 
The trace over the fermion degrees of freedom cannot be performed
analytically due to the quartic interaction term.  In order to reduce
the problem to a quadratic Hamiltonian we discretize,  
$\beta=L_\tau \Delta \tau$,
and employ the Trotter decomposition \cite{TROT}. 
The partition function can now be expressed as:
\be
Z= {\rm Tr}\, e^{-\beta \hat H} 
 = {\rm Tr}\,[e^{-\Delta \tau \hat H} ]^{L_\tau}
\approx {\rm Tr}\, 
[e^{-\Delta \tau \hat K} e^{-\Delta \tau \hat P}]^{L_\tau} ~.
\label{eq:ZZ}
\ee 
Here $\hat K$ includes the a priori quadratic pieces of $\hat H$,
the hopping and chemical potential terms, while $\hat P$ is the
on-site interaction term. In the second step, errors in measured
quantities of order $tU(\Delta \tau)^2$ are introduced.
Now that $\hat P$ has been isolated, the
discrete Hubbard-Stratonovich transformation of Hirsch \cite{HS} can 
be used to decouple the interaction.  If $U>0$,
\be
e^{-\Delta \tau U (n_{i \uparrow} - {1 \over 2})(n_{i \downarrow} -
{1 \over 2})} = \ha e^{- U \Delta \tau/4} \sum_{S_{i,\tau} = \pm 1}
e^{-\lambda \Delta \tau S_{i,\tau} (n_{i \uparrow} - n_{i \downarrow})}.
\label{eq:lamU}
\ee
Here $\cosh (\lambda \Delta \tau) = \exp(U\Delta \tau / 2)$. 
A field variable $S_{i,\tau}$ (of Ising type)
must be introduced at each lattice site and
imaginary time slice.  If $U<0$ the field $S_{i,\tau}$
couples to the charge $n_{i\uparrow}+n_{i\downarrow}$
on each site.  In either case, the end result is that the 
partition function can be rewritten as:
\be
Z=\sum_{S_{i,\tau}} {\rm Tr} \prod_{l} e^{-\hat K} e^{-\hat P'(l)} ~,
\label{eq:ZSi}
\ee
where the fermion operators appear only quadratically in the
transformed $\hat P'$ and $l$ denotes the Trotter time slice.
The interacting fermions have been replaced by
non-interacting fermions evolving in a space-- and time--dependent field.
Performing the fermion trace is now possible, and results in: 
\be
Z=\sum_{S_{i,\tau}} {\rm det}M_{\uparrow}(S_{i,\tau}) \,\,\, {\rm
det}M_{\downarrow}(S_{i,\tau}) ~,
\label{eq:Zud}
\ee 
where the detailed forms of the matrices $M_{\sigma}$, which have
dimension the number of spatial lattice sites $N$, are written down 
by Blankenbecler \etal \cite{BlnkWh}.

One crucial bottleneck in the algorithm is that while
at high temperatures the product of the fermion determinants is 
positive and can be
used as a statistical weight, at low temperatures (high $\beta$)
when $U>0$ and the system is away from
half-filling, this is no longer always true.  In this case, as 
$\beta$ increases the product can become negative.
This is referred to as the ``sign problem''.
The sign problem precludes simulations for large lattices
at low temperatures.
When $U<0$ no sign problem occurs, and this is one reason the
attractive Hubbard model is easier to study.

The Monte Carlo simulation proceeds by going through
the space--time lattice of $S_{i,\tau}$ and computing the ratio of the
product of fermion determinants before and after a single
spin is flipped.  Naively, the simulation time
scales as the fourth power of the number of sites $N$.  
One power of $N$ comes from updating all the sites.
For each such update, the calculation of the determinant is of
order $N^{3}$.  However, one can reformulate the algorithm to
scale with one lower power of $N$ as follows:
instead of evaluating the new determinant from scratch,
the ratio of the determinants
may be evaluated in 
terms of the elements of $M_{\sigma}^{-1}$, the fermion Green
function.  One then must keep track of $M_{\sigma}^{-1}$, which of course
changes as moves are accepted.  However, the update of $M_{\sigma}^{-1}$ 
resulting from a change to $M_{\sigma}$ is only an $N^{2}$ procedure, due
to the specific local nature of the change in $M_{\sigma}$.
In summary, the technique involves the manipulation
(mostly multiplication) of matrices of dimension the number
of spatial sites in the system.  This number may be 
up to several hundred on present--day computers.

The inputs to the computer code are the quantities that specify
the physical system.
They are: the spatial size of the system and
the inverse temperature $\beta$.
Also, the parameters in Hamiltonian (\ref{eq:Hh}), 
$t, U_j,$ and $\mu$, need to be specified.  
Finally, the parameters of the
simulation, the Trotter time interval $\Delta \tau$
(which then determines the number of time steps $L_\tau$), 
the number of warm--up and measurement sweeps, the time
between measurements etc. must be selected.  
Quite some care needs to be taken in
determining these latter quantities to ensure the simulation
results are meaningful.

One crucial feature of the Determinant Monte Carlo simulations
is that the fermion Green functions $G_{ij}(\tau,\tau')
=\langle T c_{i}(\tau)c_{j}^{\dagger}(\tau') \rangle$ are readily
obtainable.  
In fact, $G$ is just the inverse of the matrix $M$ whose 
determinant gives the Boltzmann weight. 
It is available ``free of charge'' (at least for $\tau = \tau'$)
in the sense that it has already been constructed in moving the field 
variables.  With the single--particle Green function in hand, we
can measure quantities involving more than two fermion operators by
performing the appropriate Wick contractions and expressing them in 
terms of products of single--particle propagators.  
There is considerable flexibility in the quantities one can calculate,
and it is possible to decompose diagrams and separate self-energy and 
vertex contributions.  
This is useful when comparing quantum simulations with
analytic calculations \cite{BICKERS}.

The Determinant Monte Carlo algorithm as described above has been 
extensively used for over a decade now. 
The more reliable applications have mainly been due to a number
of recent technical advances.
Here we only mention the most important one for our present studies:
the introduction of matrix stabilization procedures has 
significantly improved the ability to reach low temperatures 
\cite{STABLE,WhHbk}. With these techniques inverse temperatures as 
high as $\beta t = 20$ can be attained.

In the present study, we have used $U=8t$ (equal to the band width)
combined with a Trotter time
step $\Delta \tau = 1/12t$, ensuring small Trotter errors.
Experience from previous studies \cite{WhHbk}
and independent tests in the present work
teach that an inverse temperature $\beta t = 8$ is sufficiently high
to effectively yield ground-state thermodynamic and correlation functions
on the lattices considered.
This temperature is used throughout. We typically use 200 to 300
warm--up sweeps and 2000 to 4000 measurement sweeps. 
A simplifying detail of the present problem, in which
a fraction $f$ of the sites have $U=0$, is that the corresponding 
Hubbard--Stratonovich coupling $\lambda$ in (\ref{eq:lamU}) equals zero 
and the local electron Green function does not need updating.
We study square lattices with linear size 
up to $L=10$ and average over up to 40 disorder realizations.

Like the original Hubbard model, Hamiltonian (\ref{eq:Hh}) preserves a 
particle--hole symmetry at half--filling ($n=1$ and $\mu=0$), 
i.e.~it is invariant under the ``staggered'' particle--hole 
transformation $c^\dagger_{i\sigma}\rightarrow 
(-1)^{i} c^{\phantom \dagger}_{i\sigma}$.
This symmetry ensures that there is no sign problem at $n=1$.
Note that the particle--hole symmetry corresponds to different chemical 
potentials on the two constituents. 
When going off half--filling we do encounter
the sign problem. However, we find it to be less detrimental than in the
case without disorder. Aided furthermore by the disorder averaging it is
possible to obtain meaningful data 
off half--filling (see also section \ref{results}).
We further support our findings by comparison to results of DMFT 
calculations.

\subsection{Dynamical Mean--Field Theory} \label{DMFT}
For classical spin models (e.g.~the Ising model) 
it is well known that the Weiss
molecular field (or: mean field) theory becomes exact in the limit of 
high spatial dimensions. 
For lattice electrons this limit was introduced only recently;
with the proper scaling of the hopping element in (\ref{eq:Hh}),
$t = t^*/\sqrt{z}$ ($z$ is the number of nearest neighbors)
it leads to  a quantum mechanical dynamical
mean--field theory (DMFT). Here we give a brief description of this
approach to our problem at hand; more details on the method are 
found in the literature \cite{MV,UlmJan}.

Thermodynamic properties are determined from the averaged 
grand potential $\Omega_{\rm av}$:
\begin{equation}
\beta\Omega_{\rm av}  = - \langle
\big( \ln \;  {\rm Tr} \; \exp (- \beta \hat{H} ) \big) \rangle_{\rm av} ~,
\label{Freeenergy}
\end{equation}
where $\langle \dots \rangle_{\rm av}$ denotes the disorder average.
The main simplification in the limit $d\to\infty$ lies in the reduction of
the average, which has to be performed only on a single site \cite{Janis92}
(for each allowed value of $U_j$, and using a {\em coherent potential} type 
of approximation).
The explicit expression for the grand potential in
the paramagnetic phase is given by:
\begin{eqnarray}
\beta\Omega_{\rm av} &=& - N \sum_{\sigma n} \int_{- \infty}^{\infty}
\mbox{d}E \, N^0 (E)
\ln [ i \omega_n + \mu - \Sigma_{\sigma n} - E]
\nonumber \\[10pt]
& + &  N \sum_{\sigma n} \ln G_{\sigma n}^{-1} - N
\big\langle
\ln {\cal Z} \{  G, \Sigma ,  U_j \} \big\rangle_{\rm av} \, ,
\label{Omega}
\end{eqnarray}
where $\omega_n = (2n+1) \pi/\beta $ are Matsubara frequencies and 
$N^0(E)$ is the density of states (DOS) of non--interacting electrons
(scaled to be suitable for comparison with the square lattice \cite{DOS}).
In (\ref{Omega}) also appear:
the generalized atomic partition function ${\cal Z}$ (shown to be 
equivalent to that of a single--impurity Anderson model \cite{GeoJar92}), 
and the complex quantities
$G_{\sigma n}$ and $\Sigma_{\sigma n}$ which at this point enter as
variational parameters. 
The {\em physical} values of $G_{\sigma n}$ and $\Sigma_{\sigma n}$
(namely, the local Green function, 
$G_{\sigma n} \equiv G_{ii, \sigma n}$,
and the electron self-energy, 
$\Sigma_{\sigma n} \equiv \Sigma_{ii, \sigma n}$,
respectively) correspond
to those for which (\ref{Omega}) is stationary
\cite{Janis92a}.
The stationarity conditions:
$\delta \Omega_{av}/\delta G_{\sigma n} = 0 \; , \;
\delta \Omega_{av}/\delta \Sigma_{\sigma n} = 0$,
yield two coupled sets of self--consistent equations for $G_{\sigma n}$
and $\Sigma_{\sigma n}$:
\begin{eqnarray}
G_{\sigma n } & = & - \int\limits_{- \infty}^{\infty} \mbox{d}E
\frac{N^0(E)}{i \omega_n + \mu - \Sigma_{\sigma n} - E } 
\label{Dyson} \\
G_{\sigma n} & = &  \int\limits_0^\beta \mbox{d}\tau \; e^{i \omega_n \tau}
\langle \langle c^{\phantom *}_{\sigma}
 (\tau) c_{\sigma}^* (0) \rangle_T \rangle_{\rm av}
\label{Singlesite}
\end{eqnarray}
where $\langle \dots \rangle_T$ denotes the thermal average
(which depends on $G_{\sigma n}$ and $\Sigma_{\sigma n}$ implicitly).
Physically, the Dyson equation (\ref{Dyson}) describes the local 
Green function of 
independent electrons moving in a homogeneous dynamical potential 
$\Sigma_{\sigma n}$. It can be solved by a simple integration for each
Matsubara frequency. 
The functional integral (\ref{Singlesite}) on the other hand is highly 
non--trivial since it couples all Matsubara frequencies.
The interacting, disordered problem is mapped onto an ensemble of 
single--impurity problems, complemented by a self--consistency condition which
introduces the lattice into the problem \cite{DobKot,Janis92}.
We emphasize that, in contrast to conventional mean--field theories, in this
dynamical mean--field theory (DMFT), the action remains 
time dependent, i.e.~local fluctuations are retained.   

The local interacting problem (\ref{Singlesite}) is solved numerically
using  a finite--temperature, auxiliary--field
Quantum Monte Carlo (QMC) method \cite{Hirsch86}.
Quite similar as in the finite--dimensional QMC algorithm described above, 
the electron--electron interaction 
is formally replaced by an interaction of independent electrons 
with a dynamical, auxiliary field of Ising--type spins.
To this end the interval $[0,\beta]$
is again discretized into $L_{\rm \tau}$ steps of size 
$\Delta\tau=\beta/L_{\rm \tau}$.  
Equivalently, there is a high energy cut--off of Matsubara frequencies, 
i.e.~$|\omega_n|= |2n+1|\pi/\beta <\pi/\Delta\tau$, 
$n=-L_{\rm \tau}/2,\dots, L_{\rm \tau}/2-1$.
The computer time grows like $L_{\rm \tau}^3\propto\beta^3$, restricting 
$L_{\rm \tau}$ to values below $\sim 150$ and $\beta t^*\le 50\dots 70$
on present supercomputers. 
For small $L_{\rm \tau}$ $(L_{\rm \tau}\leq 20)$ one can perform a full enumeration
(instead of Monte Carlo sampling) of all $2^{L_{\rm \tau}}$ possible 
configurations of the auxiliary field.
Sign problems turn out to be absent in the DMFT approach.

Self--consistency is obtained iteratively as follows: the Green function 
$G$ (omitting indices) is calculated from some initial self energy, 
e.g.~$\Sigma\equiv 0$, by the Dyson equation (\ref{Dyson}).
Then, the new Green function $G_{\rm new}$ is determined by
solving (\ref{Singlesite}) for each value of $U_j$ with the QMC method,
and averaging. 
Finally, the calculation of the new
self energy $\Sigma_{\rm new}=\Sigma-G^{-1}_{\rm new}+G^{-1}$
completes one iteration. 
Typically $10$ iterations with 20000 MC sweeps are 
necessary to obtain a convergence of $\sim 10^{-3}$. 
Close to a phase transition the convergence is much slower
and the statistical errors are  larger due to strong fluctuations.
At large $U$-values $(U>12t)$ the Monte Carlo sampling becomes more and
more inefficient due to ``sticking'' problems, i.e.~there are two (or more)
minima in the free energy and the single spin--flip algorithm is no longer able
to transfer between them.

The self--consistent equations (\ref{Dyson}) and (\ref{Singlesite})
can easily be extended to a phase with AF long range order
(For details of the implementation see \cite{UlmJan}). One uses the property
that, in the case of finite sublattice magnetization but zero 
total magnetization, i.e.~excluding ferrimagnetism, the Green function 
for an $\uparrow-$spin on sublattice $A$ is equal to that of a 
$\downarrow-$spin on sublattice $B$ and vice versa. 

\section{Quantities of Interest}
In this paper, we only study thermodynamic quantities and 
static correlation functions. At a later stage also (imaginary-) 
time--dependent correlation functions will be of interest, e.g. to
study frequency--dependent response functions.

In the QMC calculations, the (equal--time)
electron Green function $G_{ij\sigma} =
\langle c_{i\sigma}^{\phantom \dagger}c^\dagger_{j\sigma} \rangle = M^{-1}_{ij\sigma}$,
described in section \ref{DetQMC} immediately leads to electron
densities, 
$\langle n_{i\sigma} \rangle = 1 -  G_{ii\sigma}$, 
and static correlation functions.
The magnetic correlations which give crucial information on the physics
of the Hubbard model can be monitored by
measuring the equal--time spin--spin correlation function
$ c(l)=\langle (n_{j\uparrow}-n_{j\downarrow})
(n_{j+l \, \uparrow}-n_{j+l \, \downarrow})\rangle$
and its Fourier transform $S(q)=\sum_{l}e^{iq\cdot l}c(l)$,
the magnetic structure factor. Specifically, a finite--size scaling
study of $S(\pi,\pi)$ leads to the sublattice magnetization $M$
of the phase with AF long range order, 
according to spin wave theory \cite{Huse88}:
\be
\frac{S(\pi,\pi)}{L^2} = \frac{M^2}{3} + {\cal O}(1/L) ~, \label{eq:Msw} 
\ee
where $L$ is the linear system size. 

In the DMFT calculations (see section \ref{DMFT}),
the resulting Green functions also lead to, 
for instance, the densities $n_{\sigma}$ and sublattice magnetization $M$.
The Green function $G_{ii\sigma}$ is just the Fourier transform of
$G_{\sigma n}$ in section \ref{DMFT}. Note that in this approach quantities
like the order parameter $M$ are calculated in the thermodynamic limit,
so no finite--size scaling is required.
The vanishing of $M$ is used to determine
phase boundaries as a function of parameters like $\mu$, $U$, $T$, 
or $f$, leading to e.g. the N\'{e}el temperature $T_{\rm N}$.
A different way to obtain these phase boundaries
is to calculate the corresponding susceptibility $\chi_{\rm AF}$
(via two--particle correlation functions; see \cite{UlmJan} for details).
Divergence of $\chi_{\rm AF}$ signals a continuous phase transition. 

\section{Results and Discussion} \label{results}
\subsection{Magnetic order} \label{Magnord}
First, we study the effect of an increasing concentration $f$ of $U=0$ 
sites on the stability of AF long range order at $n=1$. 
The static AF structure factor $S(\pi,\pi)$ is calculated at $U=8t$ for 
different lattice sizes and temperatures. For $L\leq 10$, $S(\pi,\pi)$ 
is found to saturate at $T\approx t/8$. 
>From the saturated values
the ground--state sublattice magnetization $M$ can be extrapolated using 
finite--size scaling, according to (\ref{eq:Msw}). 
Scaling plots for different $f$ at $U=8t$ are shown in Fig.~1. 
For $f\leq 0.36$, $S(\pi,\pi)/L^2$
extrapolates to a finite value in the thermodynamic limit.
Long range order has disappeared for $f\geq 0.5$. Fig.~2 presents the 
extrapolated values $M(f)$. For small disorder $M$ is found to 
increase with $f$, 
it reaches a maximum around $f=0.1$, and 
eventually vanishes at $f_c\approx 0.4$.
This initial increase of the order parameter
prior to the formation of a disordered phase is remarkable.
Similar phenomena have however been seen 
in studies of the Hubbard model with random site energies in
infinite dimensions  
(an initial increase with disorder of the critical temperature for 
AF order \cite{UlmJan}).
Local enhancement of AF fluctuations by vacancies has
also been reported recently \cite{DAGOTTO}.

\begin{figure}
\vskip-10mm
\hspace*{ 20mm}
\psfig{file=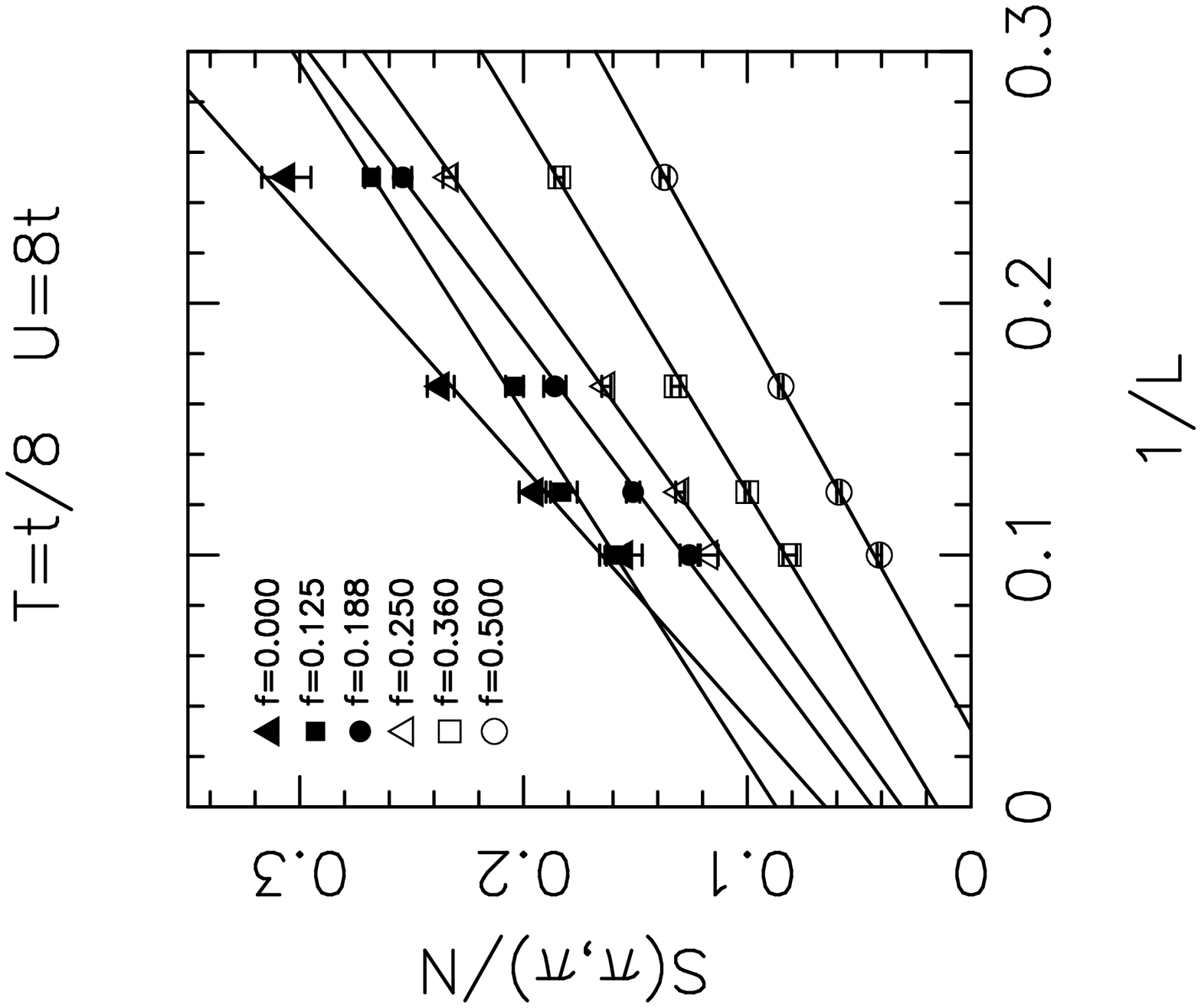,height=3.8in,width=5.4in,angle=-90}
\vskip-02mm
\caption{
Finite--size scaling of the 
antiferromagnetic structure factor $S(\pi,\pi)$ on the square lattice
at half-filling for $f=~0,~0.125,~0.188,~0.25,~0.36,~0.5$ 
(top to bottom; $f$ is the fraction of $U=0$ sites), 
computed using the Determinant QMC method with $U=8t$ and $T=t/8$.
$L$ denotes the linear system size and $N$ the number of sites 
($N=L^2$). 
For values of $f$ that do not correspond to an integer number 
of defects, we have interpolated between the results for the 
two bracketing concentrations.
}
\end{figure}

\begin{figure}
\vskip-10mm
\hspace*{ 20mm}
\psfig{file=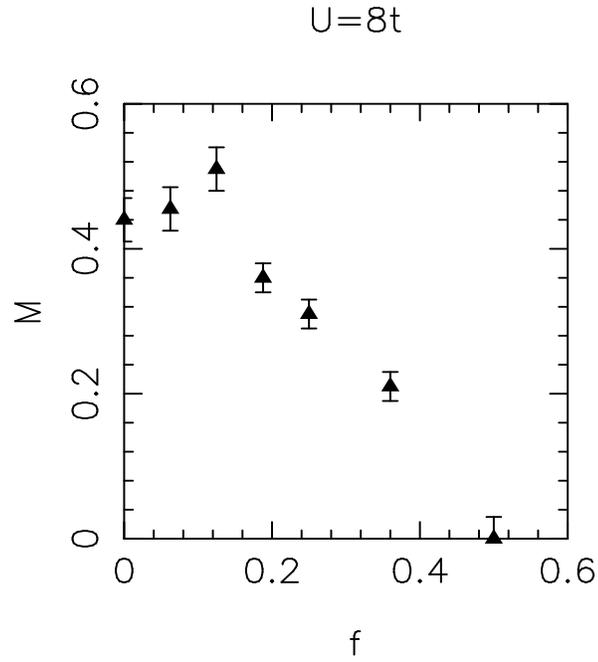,height=3.8in,width=5.4in,angle=-90}
\vskip-02mm
\caption{
Ground--state staggered magnetization $M$ as a 
function of fraction $f$ of $U=0$ sites, as extrapolated from Fig.~1.
}
\end{figure}

\begin{figure}
\vskip-15mm
\hspace*{ 20mm}
\psfig{file=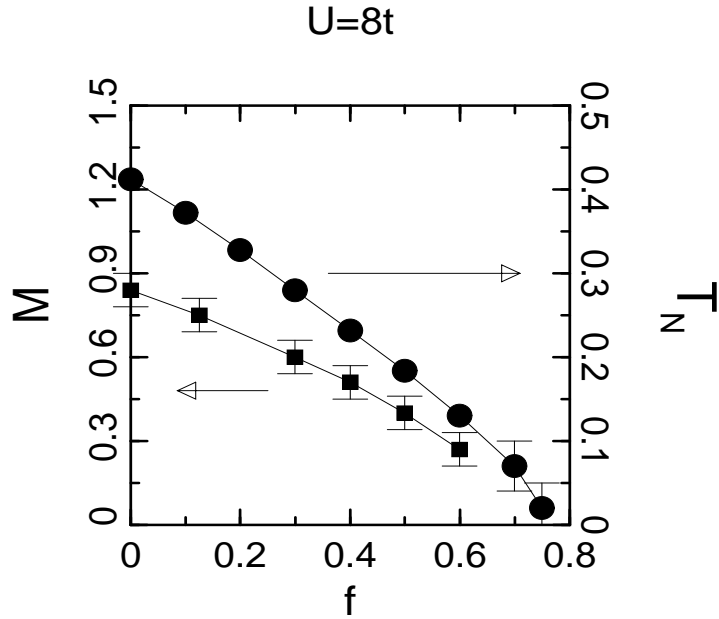,height=4.1in,width=5.4in,angle=-00}
\vskip-10mm
\caption{
Staggered magnetization $M$ and N\'{e}el
temperature $T_{\rm N}$ as a
function of fraction $f$ of $U=0$ sites, computed using Dynamical
Mean Field Theory (DMFT). $T_{\rm N}$ values are obtained by
extrapolating $\Delta\tau$ to zero (see section 2.2)
those for $M$ are not, since extrapolation is cumbersome.
Also, small values of $M$ are hard to compute.
}
\end{figure}

For the present model, at half-filling, we do not find 
initially enhanced AF order in the {\em infinite--dimension} approach
as can be seen in Fig.~3 (Due to the neglect of spatial fluctuations,
we do of course find that AF order survives longer when increasing
disorder: $f_c\approx 0.75$. This can be higher than the percolation
threshold because no sites are actually removed). 
On the other hand, also in this approach we can see that AF order is
enhanced by introducing $U=0$ sites:
we find that AF order occurs at densities for which the pure system
is not ordered. A calculated $(f,n)$ phase diagram shows that AF order
persists up to increasing values of $n$ (relative to $f=0$, for which 
in DMFT AF order occurs up to $n=1.14$, i.e. somewhat off half-filling) 
when the disorder increases from $f=0$ to about $f=0.36$ \cite{MPRG}.
Therefore, again the presence of impurity sites stabilizes the magnetic 
order.
We note that the calculation of this phase diagram using the Determinant 
QMC method in finite dimension is very hard, if not impossible, because 
of the sign problem.
We, however, expect the phase boundary between AF and disordered phase
to be of similar form (but shifted towards half--filling).

To come to an explanation for this disorder-stabilized magnetic order we 
argue that the situation is different in one important aspect from the 
case of electron-- or hole--doping of the antiferromagnet at half--filling
in the pure Hubbard model, which destroys AF order rapidly.
In the pure case, the extra electrons (or: holes)
are mobile and therefore particularly effective in destroying 
long--range order. With $U=0$ sites, however, extra particles will be 
localized on these sites, where they don't have to pay the 
on-site repulsion energy. So whereas magnetic moments are destroyed 
locally (on the $U=0$ sites),
the long--range order originating from the $U\neq 0$ sites will be able
to persist much longer.
That $U=0$ sites are not so detrimental to AF order can also be seen
from the fact that the usual perturbative argument to obtain an AF 
(superexchange) Heisenberg coupling in the pure model still goes
through with an intermediate $U=0$ site.
Finally, a qualitative way to understand the enhanced $M$ at $n=1$ comes 
from spin wave theory: normally the quantum fluctuations of spin waves 
reduce the magnetization.
The effect of disorder can be thought of as to introduce a lifetime, 
which can be seen to reduce the effect of the spin waves on $M$, 
thereby enhancing it.
Analogously, introducing $U=0$ sites into the negative$-U$ 
Hubbard model we expect to lead to enhanced  
charge--density wave order and superconductivity.
The spin--wave argument may also explain the absence of the enhancement 
of $M$ in DMFT in Fig.~3: in this approach the 
coherent--potential approximation (CPA) is used to treat disorder 
(see section \ref{DMFT}) and the CPA is known not to describe 
localization (in this case of spin waves), because it does not introduce
spatial fluctuations \cite{UlmJan}.

\subsection{Charge order} \label{charord}
A second type of order that is of interest is charge order, which 
generally is signalled by a range of constant density (or:
zero compressibility $\kappa = \partial n/\partial \mu$) when the 
chemical potential is varied. 
For instance, in the Quantum Hall effects the occurrence of
incompressible states is the central theme. In the pure Hubbard model
a {\em charge gap} (proportional to $U$) occurs around $n=1$:
the system resists against having doubly occupied sites. 
To study the effect of disorder by having $U=0$ sites, we calculate 
the (disorder-averaged) density $n$ when varying $\mu$ in (\ref{eq:Hh}) 
at $U=8t$ for a few values of impurity concentration $f$. To get a more 
detailed picture, we discriminate between $U=0$ and $U=8t$ sites. 
The results are displayed in Figs.~4 to 7.

\begin{figure}
\vskip-10mm
\hspace*{ 20mm}
\psfig{file=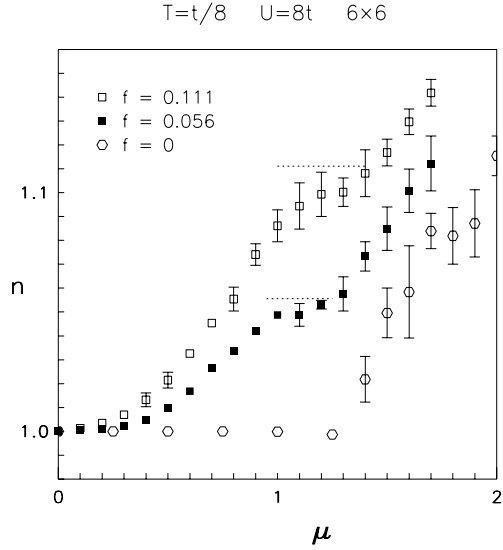,height=3.6in,width=3.5in,angle=-00}
\vskip-10mm
\caption{
Average total density $n$ as a function of $\mu$
for different impurity concentrations $f = 0,~0.056,~0.111$.
Dotted lines indicate the corresponding values $1+f$.
Calculated using
Determinant QMC on $6 \times 6$ lattices with $U=8t$ and $T=t/8$.
Error bars are within the symbol size when not shown.
}
\end{figure}

\begin{figure}
\vskip-20mm
\hspace*{ 20mm}
\psfig{file=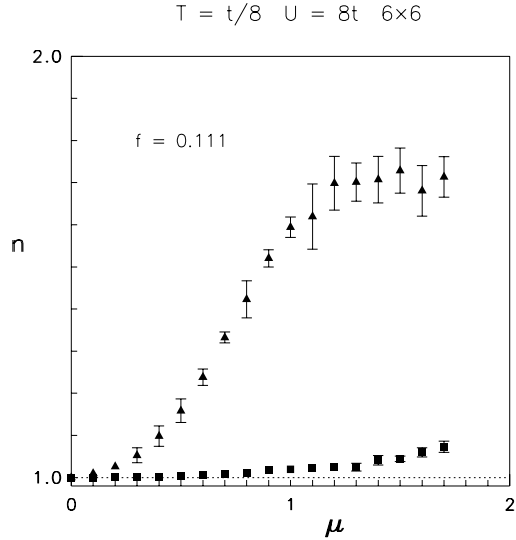,height=3.6in,width=3.5in,angle=-00}
\vskip-10mm
\caption{
Average density $n$ on $U=0$ sites (triangles) and
$U=8t$ sites (squares)
separately for impurity concentration $f=0.111$.
The saturation value for $n(U=0)$ is about 1.71.
Details as with Fig.~4.    
}
\end{figure}

\begin{figure}
\vskip-10mm
\hspace*{ 20mm}
\psfig{file=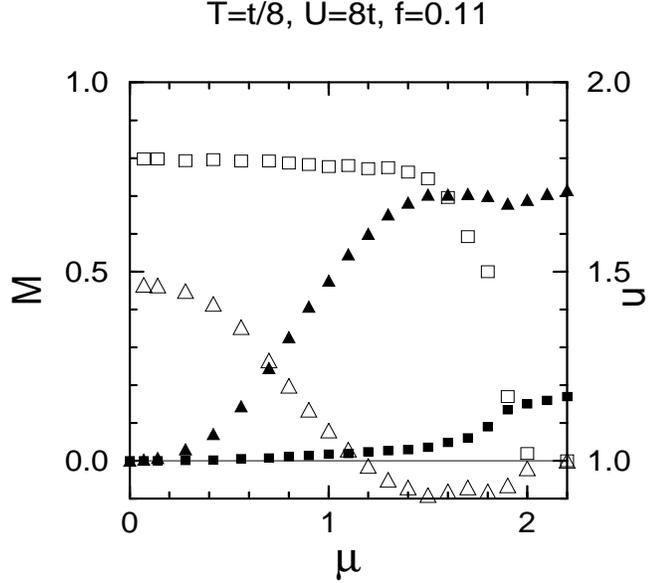,height=4.1in,width=5.4in,angle=-90}
\vskip-10mm
\caption{
Average density $n$ (filled symbols)
and sublattice magnetization $M$ (open symbols)
on $U=0$ sites (triangles) and $U=8t$ sites (squares) separately.
The saturation value for $n(U=0)$ is about 1.70.
Calculated using DMFT for impurity concentration $f=0.11$,
$U=8t$ and $T=t/8$. Fig.~5 shows the corresponding results
for $n$ for $d=2$.    
}
\end{figure}

\begin{figure}
\vskip-10mm
\hspace*{ 20mm}
\psfig{file=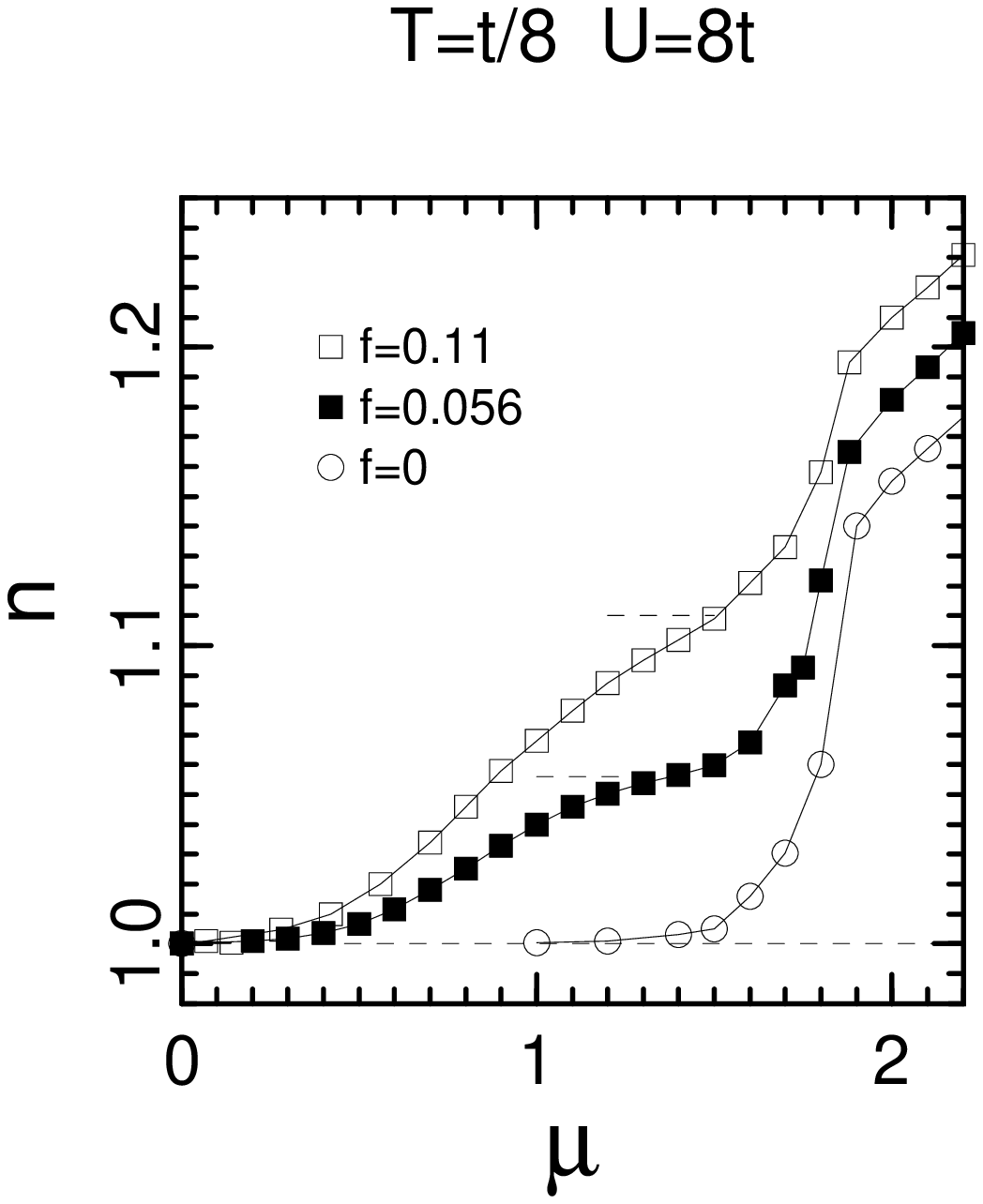,height=4.1in,width=5.4in,angle=-00}
\vskip-15mm
\caption{
Average total density $n$ as function of $\mu$
for different impurity concentrations $f = 0,~0.056,~0.11$, as
calculated using DMFT. Dashed lines indicate the occurring values $1+f$.
Details as with Fig.~6. Fig.~4 shows the corresponding results
for $d=2$.        
}
\end{figure}

Before embarking on a discussion of the results,
we first discuss the specific averaging procedure that has been used 
to obtain the result for $n(\mu)$ in Figs.~4 and 5 (using
Determinant QMC on $6\times 6$ lattices). Because of the sign problem
it is not possible to obtain reliable information at low enough
temperatures on large lattices in the pure model 
for densities $1 < n < 1.3$ (\cite{WhHbk}, see also section 
\ref{DetQMC}). Introducing disorder turns out to help us in two ways:
(i) in cases where the sign problem is less severe, e.g. for $n=1.4$, 
the computed average sign is closer to 1 in the presence of disorder, 
which means that the problem is more tractable, and (ii) the disorder 
average that one has to perform leads to a natural way of 
{\em taming} the sign problem. 
Concerning (ii): since for each realization of (quenched) 
disorder a relatively small part of phase space (but hopefully large 
enough to be representative) is sampled, it can happen that for some
realizations one is {\em hit} by the
sign problem, but for others one is not. 
This can be seen from the average sign
which becomes very small, or from calculated quantities which occasionally
deviate a lot from the majority of results (or even become unphysical, e.g. 
negative densities).  To include such {\em outliers} in an average clearly
makes no sense. A more reliable and robust way of determining the disorder
average, i.e. one which is not very sensitive to the type of distribution 
which governs the outcomes, is the so-called {\em midmean}, or: 
$25 \%$ trimmed mean (an average over the middle half of the ordered 
data set \cite{midmean}).
We have used this mean for the $d=2$ QMC calculations; using another 
robust mean, the broadened median \cite{midmean}, gives very similar 
results. We like to stress that this approach is not limited to the
disordered case; in fact, we have applied it to the pure ($f=0$) model
off half-filling as well (Fig.~4) by having different starting 
configurations take the role of different disorder realizations.

In Fig.~4, it is possible to discern for non-zero $f$
two flat regions in $n(\mu)$, i.e. two hardly compressible states. The
first still occurs for $n=1$, but has a much reduced charge gap, whereas
a second gap develops at a density somewhat below $n = 1+f$. One would 
expect a plateau exactly at $n=1+f$ if the $U=0$ sites would fill 
independently of the $U \neq 0$ sites (because above $1+f$ the $U \neq 0$ 
sites would have to start to be doubly occupied). 
Since the density at $U=0$ sites does not saturate
at 2, but at a lower value ($n=1.76$ and 1.71 for $f=0.056$ and 0.111, 
respectively; see Fig.~5), clearly the two sets of sites are coupled.
On the other hand, one notes from Fig.~4 that the 
second plateau for non--zero $f$ terminates almost exactly at the 
$\mu_{\rm c}$ for $f=0$. 
Therefore, the chemical potential needed to force double
occupation of the $U \neq 0$ sites is unaffected by the presence of
$U=0$ impurities. The fact that $n$ at the $U \neq 0$ sites remains
pinned at 1 (Fig.~5) up to about the same value of $\mu$ 
as for $f=0$ confirms this.
As discussed above, indeed the error bars for non--zero $f$ are smaller 
than for the pure case, indicating a less--severe sign problem.

The remnant gap at $n=1$ is in fact the more remarkable one,
since the gap at $n=1+f$ is a clear descendant of the one at $n=1$
in the pure model (see above). This can also be appreciated from 
considering our model first with $t=0$; for that case $n(\mu)$
will have two plateaus, at $n=1+f$ and $n=1-f$ (not mentioned above, 
but present because of particle-hole symmetry). Turning on the hopping
not only leads to a rounding of the sharp steps, but also to the
appearance of a new plateau at $n=1$. 
We argue the gap at $n=1$ to be due to {\em induced} AF order 
on the $U=0$ sites. 
To support this we have computed within DMFT for $f=0.11$ 
besides the densities also the sublattice magnetizations $M$ (Fig.~6).
The decrease of $M$ is clearly coupled to the pinning of $n$ at 1, both
for $U=0$ and $U=8t$ sites. For small $\mu$ there is AF order on both 
types of sites, whereas for $\mu > 1.6$ the density on $U=0$ sites 
saturates, the density on $U=8t$ sites starts to increase, and AF order 
disappears. Therefore, certainly for this rather large value of $U$, 
both charge gaps turn out to be intimately linked to AF magnetic order.

Somewhat surprisingly, $M$ on the $U=0$ sites becomes negative when $n$ 
on these sites saturates. The reason is that electrons at $U=0$ sites are
more strongly localized when their spin is parallel to the neighbors than
when it is antiparallel (because of Pauli's principle). Therefore, the
net moment on $U=0$ sites is parallel to its neighbors, i.e. opposite to
the total staggered magnetization.

For completeness we also show the total density as function of $\mu$
computed using DMFT in the thermodynamic limit (Fig.~7). The overall 
picture of the $d=2$ QMC calculations on $6\times 6$ lattices
(Fig.~4) is confirmed, although
the plateaus close to $1+f$ are less well developed, certainly for $f=0.11$.
Another difference is that things happen at a slightly higher
value of $\mu$, because fewer fluctuations are accounted for in DMFT.

Finally, we strengthen our case for an incompressible state close to
$n=1+f$ by showing in Fig.~8 the kinetic energy (as computed on a 
$4\times 4$ lattice) as function of $n$. Clearly there is a dip in the
electron mobility at $n=1+f$. A similar argument, using cusps in the 
kinetic energy as evidence for incompressible states, was found to work
for the boson Hubbard model \cite{boshub}.
We further note that additional support can be found
in calculations of the density of states within DMFT 
that we present elsewhere \cite{MPRG}.

\begin{figure}
\vskip-05mm
\hspace*{ 20mm}
\psfig{file=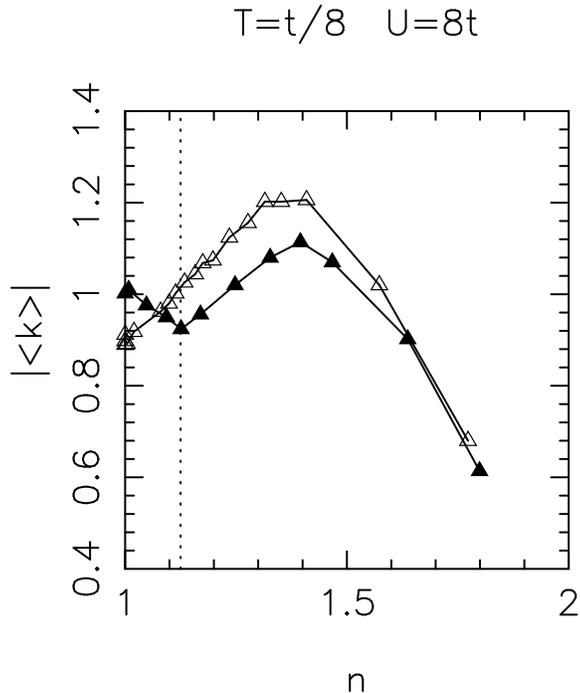,height=4.0in,width=5.4in,angle=-90}
\vskip-02mm
\caption{
Kinetic energy $|\langle K \rangle|$ as function
of density $n$ for impurity concentrations $f = 0$ (open triangles)
and $f=0.11$ (filled triangles).
Calculated using Determinant QMC on $4 \times 4$ lattices with
$U=8t$ and $T=t/8$. 
}
\end{figure}

In summary, using Determinant QMC calculations on finite lattices for $d=2$
combined with the infinite-dimension approach in the thermodynamic limit, 
we have studied the effect on ordered states in the Hubbard model of 
introducing {\em disorder of the third kind}, namely
in the on-site {\em interaction}. We show that turning 
off the on-site interaction on a fraction of the sites of the lattice
may enhance and stabilize AF order. It also leads to the occurrence of two
incompressible states instead of one in the case without disorder; 
closer scrutiny of density and sublattice magnetization on the two types
of sites teaches that both states are associated with AF order.

\newpage
\section*{Acknowledgments}
This paper is dedicated to Hans van Leeuwen on the occasion of his
65th birthday. The first author thanks Hans for introducing him to
the wonderful world of the Hubbard model and for encouragement,
guidance, and collaboration during several years of studies of this model.
He furthermore is grateful to the UC Davis Physics Department for 
hospitality during the period in which the presented work originated.
The research described here is supported in part by NSF-DMR-95-28535.


\begin{thebibliography}{99}
\bibitem{DOS}
We choose a semi--elliptic DOS, 
$N^0(E) = [(2t^*)^2 - E^2 ]^{1/2}/(2 \pi t^{*2})$.
For comparing with the square lattice $(z=4)$, we use $t^*=2t$ and 
hence the DMFT calculations use the same bandwidth $(W=8t)$
and second moment $(\langle E^2\rangle=4t^2)$ of $N^0(E)$ 
as in $d=2$.
DMFT results are fairly insensitive to the choice of DOS.
\end{thebibliography}
\end{document}